\documentclass[12pt]{JHEP}

\usepackage{epsfig}
\usepackage{amsmath}
\usepackage{amssymb,amsfonts}

\advance\hoffset by -.35in
\newcommand{\ar}[2]{\left[ {}^{#1}_{#2} \right]}

\preprint{hep-th/0205073\\DAMTP-2002-50}

\title{\Large\bf Instantons corrections to the effective action of the CHL
string and its type I dual}

\author{P. Bain\\

\vskip 24pt

Department of Applied Mathematics and Theoretical Physics,\\
University of Cambridge,\\
Wilberforce Road, CB3 0WA Cambridge, U.K.\\}

\abstract{In this letter, we investigate corrections to quartic gauge
couplings in compactifications of string theories on a 2-torus without 
vector structure. First, we calculate the threshold corrections to
$F^4$ terms in the heterotic CHL string. Then, using the CHL
string/type I duality dictionary, we map these corrections to
perturbative and non-perturbative effects in type I string compactified
on 2-torus without vector structure. Comparing the perturbative
terms provides a quantitative test of the S-duality conjecture. The
non-perturbative contributions are due to D-strings wrapping the
torus. A T-duality along one of the 
compactified directions allows us to compare these instanton
corrections to the ones obtained in a standard type I
compactifications. The most striking feature is that these
non-perturbative couplings turn out to be identical for both models.}

\begin{document}

\section{Introduction}

A fruitful way to strengthen a non-perturbative conjectured duality
that relates two different string theories is to compare the 
higher derivative terms which appear in the effective actions of both
theories. In the context of the type I/heterotic $SO(32)$ equivalence
\cite{w1,pw}, 
this programme has been first settled in \cite{bk,bfkov,ko,b1} for a
special class of corrections which are half-BPS satured, and
therefore, can receive corrections only from half-BPS states. On the
heterotic side, these terms are related to anomaly cancelling
couplings and appear only at one-loop, for compactifications on torii
$T^d$ with $d<6$. Indeed, in this situation, the only half-BPS
instantons are due to fundamental string world-sheets wrapped on the
target space torus.  Translated into a type I language, these terms 
become tree-level, perturbative and non-perturbative corrections
due to wrapped D-strings. In particular,
the one-loop open string amplitudes reproduce the complex moduli
dependent term of the heterotic ${\rm Tr}(F^4)$ couplings. These
calculations have been extended to configurations with Wilson lines
and/or to other half-BPS saturated couplings by
\cite{fs,gut1,ls,lsw,kop}. 

In this article, we will be interested into the model considered in
\cite{gut1}, namely compactification of heterotic string theory on a
two-torus with a Wilson line along one circle which breaks the
symmetry group $SO(32)$ to $SO(16)\times SO(16)$. After S-duality,
this model has a natural description is term of the type ${\rm
I}^\prime$ string, obtained after a T-duality along the circle with
the Wilson line. Indeed, as we will describe in details in this paper,
in this picture, the R-R tapdoles are not only globally canceled, but
also locally and the dilaton is an arbitrary constant. On the
heterotic side, part of the one-loop threshold corrections - the
complex modulus independent terms - to the
$F^4$ and $R^4$ have been obtained in \cite{gut1} and interpreted as
the effect of euclidian D0-particles in the dual type ${\rm
I}^\prime$ theory. Here, we will review this result and also calculate
the complex modulus dependent corrections for the ${\rm Tr}(F^4)$
couplings. The mapping of these terms to perturbative corrections on
the type ${\rm I}^\prime$ side will provide a quantitative test of the
duality. 

One of the aims of \cite{gut1} was to propose a way to understand the
D-instantonic contributions from the matrix model
point-of-view which describes D0-branes on top of an O8-plane and
sixteen D8-branes \cite{df,bss,bgl,bgs}. Indeed, in the context of the maximally
supersymmetric type IIB string, the 
non-perturbative contributions to the $R^4$ term \cite{gg1} have been
successfully interpreted \cite{gg2,mns,kv} from the matrix theory
point-of-view 
\cite{bfss}~: these corrections, including the numerical factor,
are reproduced by the calculation of the partition function for the
matrix model of D0-branes. However, in less supersymmetric cases
such as in type I string theory or on the effective action of
D3-branes \cite{bbg}, the relation remains elusive for far. 

Therefore, in order to understand the rules of the instanton
calculus, it is useful to have at disposal other 
examples in which D-instanton corrections appear. A particularly
interesting case, because of the relations it shares with the standard
type ${\rm I}^\prime$ string, is the compactification of type I string
theory on a torus without vector structure
\cite{bps,bianchi,witten}. T-dualizing along one direction leads to a
compactification with a O8-plane and D8-branes and locally canceled
R-R charge, on a skew torus. 
This theory is believed to have the CHL string \cite{chl} as S-dual
 which can also be seen as a compactification of ${\cal
M}$-theory on a M{\oe}bius strip \cite{dp}. The aim of the present
article is 
to calculate the one-loop threshold corrections to the ${\rm Tr}(F^4)$
term in the CHL string and to relate them to perturbative and
non-perturbative contributions on the type I side. 

In the next section, we will review the calculation of \cite{gut1} and
extend it to the complex modulus dependent corrections, allowing us to
test the duality relation in the presence of the Wilson line that
breaks the 
gauge group to $SO(16)\times SO(16)$. Then, in section 3, we describe
the CHL string construction and calculate the one-loop four-point
function for four gauge fields. Using the duality map between CHL
string and type I compactification without vector structure, we give
the intepretation of these couplings in a compactification of string
theory with O8-planes and D8-branes. In particular, we notice that the
non-perturbative corrections are {\it identical} -~up to an obvious
substitution in the action of the euclidean D-particles~- to the
contributions obtained 
in the standard type ${\rm I}^\prime$ theory in section 2. 
Finally, comparing the $U$-modulus dependent terms to one-loop
corrections on the type I side put the conjectured duality on a
firmer ground. 

We have collected, for the reader's convenience, some useful theta
functions identities in appendix A and world-sheets integrations in
appendix B.

\section{Review of heterotic thresholds and of their S-dual interpretation} 

\subsection{Heterotic thresholds in the presence of a Wilson line}

We consider the $SO(16)\times SO(16)$ heterotic string theory
compactified on a square 2-torus, which can be seen as the product of
two circles of radius $R_1$ and $R_2$. The initial $SO(32)$ gauge
group has been 
broken to $SO(16)\times SO(16)$ by putting a Wilson line $Y=( 0^8,
(1/2)^8 )$ on one of the two circles of the torus, say the first. We
also index the two gauge groups by a Greek letter~: $\alpha =1, 2$. 

We will calculate the one-loop threshold corrections to the ${\rm
Tr}_\alpha(F^4)$ using the gauging procedure \cite{ejm}. To do this,
we switch on a 
background gauge field in the Cartan sub-algebra of the gauge group~:
$F^I_\alpha = v_\alpha^I$.  In this background, the one-loop partition
function reads~: 
\begin{eqnarray*} 
Z(v)=- \frac{Q(\bar{\tau})}{4 \tau^4_2}\sum_{a,b=0,1}
\frac{\prod_{I=1}^{8}
{\vartheta}\left[{}^{a}_{b}\right](v_1^I|\tau)
{\vartheta}\left[{}^{a+n^1}_{b+m^2}\right](v_2^I|\tau)}
{\eta^{24}(\tau)}
\sum_{m^i, n^i \in {\mathbb Z}} 
\Gamma^{T, U}_{2,2}\ar{n^1\;\;n^2}{m^1\;m^2}(\tau, \bar{\tau})
\end{eqnarray*} 
where we have defined the contributions of the right-moving modes~:
\begin{equation}
Q(\bar{\tau})=\sum_{\bar{a},\bar{b}=0,1}
(-)^{\bar{a}+\bar{b}+\bar{a}\bar{b}}
\frac{{\vartheta}^4\left[{}^{\bar{a}}_{\bar{b}}\right](0|\bar{\tau})}
{\eta^4(\bar{\tau})}
\end{equation}
We have also introduced the following notation 
\begin{eqnarray*}
Z_{2,2}(T,U,\tau, \bar{\tau}) &\equiv& \sum_{m^i, n^i \in {\mathbb Z}}
\Gamma^{T, U}_{2,2}\ar{n^1 \; n^2}{m^1 m^2}(\tau, \bar{\tau})
\nonumber \\
&=& T_2 \sum_{m^i, n^i  \in {\mathbb Z}} {\rm
e}^{-\frac{\pi}{\tau_2}\frac{T_2}{U_2} \vert m^1 + n^1 \tau + (m^2  +
n^2 \tau) U \vert^2 + 2i\pi T_1(m^1
n^2 - m^2 n^1)}
\end{eqnarray*}
to define the lagrangian
representation of the lattice sum
over the Kaluza-Klein momenta and windings of the closed string. 
The indices $T$ and $U$ in $\Gamma^{T, U}_{2,2}$ will be omitted when
there is no ambiguity.  

The one-loop four-point function with four gauge fields is a half-BPS
saturated amplitude; therefore, it can receive contributions only from
half-BPS states of the heterotic string, which, in eight dimensions,
correspond to fundamental string states. 
The right-moving part of the
amplitude provides the kinematic structure, namely the well-known
tensor $t_8$ which contracts the Lorentz indices of the gauge fields.
The terms whose gauge structure is ${\rm Tr}_\alpha(F^4)$ are
obtained by taking derivatives with respect to appropriate
$v_\alpha^I$ \cite{kop,ejm}. Hence, the one-loop thresholds are given  by
\begin{eqnarray*}
{\cal I}^{\rm het}_{{\rm Tr}(F^4)} &=& \Delta_\alpha^{\rm
het}(T,U)\; t_8 {{\rm Tr}_\alpha(F^4)},  
\end{eqnarray*}
with
\begin{eqnarray*}
\Delta_\alpha^{\rm het}(T,U) &=& - \frac{{\cal N}}{3\;2^5}
\int_{\cal F} \frac{d^2 \tau}{\tau^2_2} 
\left(\partial^{(4)}_{v_\alpha^1} - 3 \partial^{(2)}_{v_\alpha^1}
\partial^{(2)}_{v_\alpha^2}  
\right) Z(v) \vert_{v=0} \equiv - \frac{{\cal N}}{3\;2^5} \int_{\cal
F} \frac{d^2 \tau}{\tau^2_2} \; \Xi(\tau), 
\end{eqnarray*}
and
\begin{eqnarray*}
\Xi(\tau) &=&  \sum_{a,b=0,1}
\frac{{\vartheta}^{8}\left[{}^{a}_{b}\right]{\vartheta}^{8}\left[{}^{a+n^1}_{b+m^2}\right]}{\eta^{24}}
\left(\frac{\vartheta^{''''}\left[{}^{a}_{b}\right]}{\vartheta\left[{}^{a}_{b}\right]} - 3 \left(\frac{\vartheta^{''}\left[{}^{a}_{b}\right]}{\vartheta\left[{}^{a}_{b}\right]}\right)^2
\right) (\tau)
\sum_{m^i, n^i \in {\mathbb Z}} \Gamma_{2,2}\ar{n^1 \; n^2}{m^1 
m^2}(\tau, \bar{\tau}) \nonumber \\
&=& 2^5\sum_{m^i, n^i \in {\mathbb Z}} \left( 3 \; \Gamma_{2,2}\ar{2 n^1 \; 
n^2}{2 m^1 m^2} - \Gamma_{2,2}\ar{2 n^1 \;\;\;\;\;\;\; n^2}{2 m^1+1 \; m^2} - 
\Gamma_{2,2}\ar{2 n^1+1 \; n^2}{2 m^1 \;\;\;\; m^2} - 
\Gamma_{2,2}\ar{2 n^1+1 \;\; n^2}{2 m^1+1 \; m^2} \right)(\tau,
\bar{\tau}) \nonumber \\ 
&=&
2^5\sum_{m^i, n^i \in {\mathbb Z}}
\left(
2 \; \Gamma^{2T,U/2}_{2,2} - \Gamma^{T,U}_{2,2}
\right) \ar{n^1 \; n^2}{m^1 m^2}(\tau, \bar{\tau})
\end{eqnarray*}
where we have used in the second line the theta function identities
(\ref{derivatives}) 
given in 
the appendix to 
simplify the expression. In the last line, we have reintroduced
explicitly the dependence on the moduli of the torus. 
The normalization constant is given by
\begin{eqnarray*}
{\cal N} = \frac{V^{(8)}}{2^8 \pi^4}
\end{eqnarray*}
where $V^{(8)}$ is a regularizing volume. We have also chosen
$\alpha^\prime=1$. 

The modular
integral can be evaluated easily using the method of orbits
\cite{dkl}. For convenience, the result is given in the appendix. 
The use of formula (\ref{clint})
gives 
\begin{eqnarray}
\Delta_\alpha^{\rm het}(T,U) &=& \frac{{\cal N}}{3} 
\left( 
2 {\rm log}\vert \eta(2T) \vert^4 - {\rm log}\vert \eta(T) \vert^4 + 2
{\rm log}\vert \eta(U/2) \vert^4 - {\rm log}\vert \eta(U) \vert^4
\right),
\label{threshet}
\end{eqnarray}
up to an infrared divergence due to the contributions of massless
states circulating into the loop. 
In the following section, we will briefly review  the type I
interpretation of these couplings 
given in \cite{gut1}. Moreover, we will see how the $U$-dependent
terms 
are reproduced by a one-loop calculation on the type I side.

\subsection{Type I interpretation}

The K\"ahler and complex  structure moduli of the square torus are
given by  
\begin{eqnarray*}
T = B_{NSNS} + i {R_1 R_2}, \qquad U =
i\frac{R_2}{R_1}.   
\end{eqnarray*}
Heterotic/type I duality tells us that the coupling constants, the
metric and the $B$ fields are related by
\begin{eqnarray*}
\lambda_s^{\rm het} = 1/\lambda_s^{{\rm I}}, 
\qquad  
g_{\mu\nu}^{\rm het} = g_{\mu\nu}^{{\rm I}}/\lambda_s^{{\rm I}}, 
\qquad 
B_{NSNS}^{{\rm het}} = B_{RR}^{{\rm I}}.
\end{eqnarray*}
We will also consider a T-dual version of this type I string theory,
which in the case of the CHL string, will provides a simple, geometrical
explanation of the reduction of the group rank. Therefore, we T-dualize
the type I string along the direction $x^1$ to obtain the type I'
theory whose compactification radii and coupling constant are related
to the type I ones as
\begin{eqnarray*}
R_1^{{\rm I}^\prime} = \frac{1}{R_1^{{\rm I}}}, 
\qquad R_2^{{\rm I}^\prime} = R_2^{{\rm I}}, 
\qquad \lambda_s^{{\rm I}^\prime} = 
\frac{\lambda_s^{{\rm I}}}{R^{{\rm I}}_1} 
\end{eqnarray*} 
The $B_{RR}$ field is mapped to the value of the RR 1-form, $A_{RR}$,
along the direction $x^2$. 
Therefore, the heterotic moduli are identified to 
\begin{eqnarray}
T=A_{RR} + i \frac{R^{{\rm I}^\prime}_2}{\lambda^{{\rm I}^\prime}_s},
\qquad U = i 
{R^{{\rm I}^\prime}_1 R^{{\rm I}^\prime}_2}
\label{hetonemap}
\end{eqnarray} 
Finally, the orientifold operator $\Omega$,
which exchanges the left and right moving sectors of closed strings 
and whose action on the type IIB string theory gives the type I string, 
is transformed to a combinaison of $\Omega$ and of the reflexion along
the T-dualized direction~: $\Omega^\prime \equiv \Omega {\cal
R}_1$. Under this T-duality, the O9-plane splits into two O8-planes,
sitting at 
opposite fixed planes ($x^1=0$ and $x^1=\pi R^{{\rm I}^\prime}_1$) under the
orientifold projection $\Omega^\prime$. The sixteen D9-branes become
D8-branes, whose positions along $x^1$ depend on their initial Wilson
lines. The $SO(16)\times SO(16)$ point corresponds to a configuration
where half of these D8-branes are sitting at $x^1=0$ on top of one of
the O8-planes and the other half at $x^1=\pi R^{{\rm I}^\prime}_1$, on
the other 
O8-plane. We also point out that this configuration is the only one
where the tadpole of the R-R 9-form is locally cancelled and the
dilaton is constant.  

Using the duality map (\ref{hetonemap}), we can expand at small
coupling and finite $R_2^{{\rm I}^\prime}$ radius the $T$-dependent
part of the  
threshold corrections (\ref{threshet})
\begin{eqnarray}
\Delta_\alpha^{\rm het}(T) &=& \frac{{\cal N}}{3} 
\left(
-\pi T_2 + 4 {\rm Re} \sum_N\sum_{m|N} \frac{1}{m}{\rm e}^{4i
\pi N T} - 2{\rm Re}
\sum_N \sum_{m|N} \frac{1}{m}{\rm e}^{2i \pi N T}  
\right)
\label{hetnp}
\end{eqnarray}
The linear term comes from the contributions of the Born-Infeld action
which describes the low-energy dynamics of the D8-branes wrapped on the
direction $x^2$ of the torus. 

The exponentially vanishing terms should be attributed to 
contributions of D0-branes whose euclidian trajectories wrap the cycle
$c_2$ of the 2-torus. However, the derivation of these couplings from
first principles, for example from the matrix model which describes the
dynamics of D0-branes in this theory \cite{df,bss,bgl,bgs}, is still
lacking.  

The non-trivial test of heterotic/type I duality that we can perform
here is to compare the $U$-dependent terms of (\ref{threshet}) to
perturbative corrections on type I side. On the type {I'} side,
they will correspond to one-loop threshold corrections due to
euclidian 
fundamental strings stretched between the D8-branes and wrapped on the
direction $x^2$. To
calculate these contributions, we use the background field method
introduced in \cite{bp,bf}. We will label the D-branes with
the couple of Chan-Paton indices $(\alpha, i)$, for $\alpha=1, 2$ and
$i=1, \ldots, 16$, according to the $SO(16)$ group they realize.  
We switch on a constant
background gauge field, for example along the
directions  $x^8, x^9$~: $F_{89} = Q B$ where $Q$ is a
generator of the gauge group.  

The oscillator frequencies of the complex coordinate
$X^8+i X^9$ of a open string ending on the D-branes labelled by the
indices $(\alpha,i)$ and $(\beta,j)$ are shifted by an amount
$\epsilon$, where  
\begin{eqnarray*}
\epsilon = {\rm arctan}(\pi q^\alpha_i B) + {\rm arctan}(\pi q^\beta_j B).
\end{eqnarray*}
$q^\alpha_i$ and $q^\beta_j$ are the eigenvalues of the gauge-group generator
acting on the Chan-Paton factors at the endpoints of the string. 
The one-loop open string partition function,
namely, the sum of annulus and of the M{\oe}bius strip diagrams, is
modified in the presence of this background
\begin{eqnarray*}
{\cal A}\left(B\right) &=& \frac{i V^{(8)}}{3\;2^{13}\pi^4}
\int^\infty_0 \frac{dt}{t}
\frac{1}{\left(2\pi^2 t\right)^4}
\frac{\vartheta_1^\prime\left(0|\frac{it}{2}\right)}
{\vartheta_1\left(\frac{i\epsilon t}{2}|\frac{it}{2}\right)}
\frac{i}{2}\left(q^\alpha_i+q^\beta_j\right)Bt
\sum_{{}^{m_1 \in {\mathbb Z}+a^\alpha+a^\beta}_{m_2 \in {\mathbb Z}}} 
\Gamma_2[m_1 \; m_2]\left(\frac{it}{2}\right)
\nonumber\\
&&~~~~~~~\times\sum_{a,b=0,1}
\frac{1}{2}\left(-\right)^{a+b+ab} 
\frac{\vartheta^3\left[{}^{a}_{b}\right]\left(0|\frac{it}{2}\right)
\vartheta\left[{}^{a}_{b}\right]\left(\frac{i\epsilon
t}{2}|\frac{it}{2}\right)}{\eta^{12}\left(\frac{it}{2}\right)} , 
\nonumber\\
{\cal M}\left(B\right) &=& -\frac{i V^{(8)}}{3\;2^{13}\pi^4}
\int^\infty_0 \frac{dt}{t}
\frac{1}{\left(2\pi^2 t\right)^4}
\frac{\vartheta_1^\prime\left(0|\frac{it+1}{2}\right)}
{\vartheta_1\left(\frac{i\epsilon t}{2}|\frac{it+1}{2}\right)} 
\;\left(i q^\alpha_iBt\right)
\sum_{{}^{m_1 \in {\mathbb Z}+2a^\alpha}_{m_2
\in {\mathbb Z}}} \Gamma_2[m_1 \; m_2]\left(\frac{it}{2}\right)
\nonumber\\
&&~~~~~~~\times \sum_{a,b=0,1}
\frac{1}{2}\left(-\right)^{a+b+ab} 
\frac{\vartheta^3\left[{}^{a}_{b}\right]\left(0|\frac{it+1}{2}\right)
\vartheta\left[{}^{a}_{b}\right]\left(\frac{i\epsilon
t}{2}|\frac{it+1}{2}\right)}{\eta^{12}\left(\frac{it+1}{2}\right)}
\end{eqnarray*}
In these expressions, we have defined the 
lattice sum on the Kaluza-Klein momenta $m_1$, $m_2$ of the open string~:
\begin{equation}
\Gamma_2[m_1 \; m_2]\left(\frac{it}{2}\right) = {\rm e}^{-\frac{\pi t}{T_2 
U_2}\left(m_1^2 |U|^2 - 2 m_1 m_2 U_1 + m_2^2 \right)}
\end{equation}
To extract the ${\rm Tr}(F^4)$, we expand these expressions to quartic
order in $B$~:
\begin{eqnarray*}
{\cal A}\vert_{B^4} &=& -\frac{V^{(8)}}{3\;2^{12}\pi^4}
(q^\alpha_i + q^\beta_j)^4 B^4
\int_0^\infty \frac{dt}{t} 
\sum_{{}^{m_1 \in {\mathbb Z}+a^\alpha+a^\beta}_{m_2 \in {\mathbb Z}}}  
\Gamma_2[m_1 \; m_2]\left(\frac{it}{2}\right), \nonumber\\
{\cal M}\vert_{B^4} &=& \frac{V^{(8)}}{3\;2^{12}\pi^4}
(2q^\alpha_i)^4 B^4 
\int_0^\infty \frac{dt}{t} 
\sum_{{}^{m_1 \in {\mathbb Z}+2a^\alpha}_{m_2 \in {\mathbb Z}}} 
\Gamma_2[m_1 \; m_2]\left(\frac{it}{2}\right).
\end{eqnarray*}
Selecting the term which corresponds to the trace structure ${\rm
Tr}(F^4)$ in these amplitudes leads us to the following threshold
corrections 
\begin{eqnarray*}
{\cal I}^{\rm I, 1-loop}_{{\rm Tr}(F^4)} = \Delta^{\rm I}_\alpha(U) \; t_8{\rm
Tr}_\alpha(F^4)  
\end{eqnarray*}
with
\begin{eqnarray*}
\Delta^{\rm I}_\alpha(U) &=& -\frac{V^{(8)}}{3\;2^{12}\pi^4}
\int _0^\infty \frac{dt}{t}  
\Bigl(2 \times 16
\sum_{{}^{m_1 \in {\mathbb Z}+a^\alpha}_{m_2 \in {\mathbb Z}}}
\Gamma_2[m_2, \; m_2] +  2  \times 16 
\sum_{{}^{m_1 \in {\mathbb Z}+a^\alpha+\frac{1}{2}}_{m_2 \in {\mathbb Z}}} 
\Gamma_2[m_1, \; m_2] \\
&&~~~~~~~~~~~~~~~~~~~~~~~~~~~~~~~~~~~ -  
2^4\sum_{{}^{m_1 \in {\mathbb Z}}_{m_2 \in {\mathbb Z}}} 
\Gamma_2[m_1, \; m_2]\Bigr)\left(\frac{it}{2}\right)
\end{eqnarray*}
These integrals are given in the appendix and the result
exactly reproduces the $U$-dependent part of (\ref{threshet}).

\section{One-loop ${\rm Tr}(F^4)$ couplings in the CHL string}
 
The $SO(16)$ CHL string can be seen as a non-abelian orbifold of the
$SO(32)$ heterotic string theory. If we decompose the 32 world-sheet
fermions which realize the affine algebra into two groups,
$\chi^I_\alpha$, $\alpha=1,2$, the action of the orbifold reads~:
\begin{eqnarray*} 
&&g_1 : X_1 \rightarrow X_1 + \pi R_1, \qquad  \chi^I_1 \rightarrow
\chi^I_1, \qquad  \chi^I_2 \rightarrow -\chi^I_2 \\
&&g_2 : X_2 \rightarrow X_2 + \pi R_2, \qquad  \chi^I_1 \rightarrow
\chi^I_2, \qquad  \chi^I_2 \rightarrow \chi^I_1 
\end{eqnarray*}
The first generator corresponds to a Wilson line $Y=(0^8, (1/2)^8)$
which breaks the gauge group $SO(32)$ 
into $SO(16)\times SO(16)$ as in the model of the previous section
while the second operator projects on the diagonal $SO(16)$ gauge
group.  

The orbifold partition function \cite{bianchi,bgmn} can be written as
the sum of the 
contributions of the different sectors of the orbifold. 
As above, we will use the background field method to
calculate the amplitude for four gauge fields. Therefore, we switch on
a background gauge field in the Cartan torus of
the diagonal gauge group. The diagonal gauge field couples to the
world-sheet fields as~: 
\begin{eqnarray*}
{\cal S}_{ws} &=& \frac{1}{2\pi}\int d^2\sigma  \Bigl( \partial X^\mu
\bar{\partial} X_\mu + B_{\mu\nu}\partial X^\mu
\bar{\partial} X^\nu + \psi^\mu \partial \psi_\mu  \\ 
&&~~~~~~~~~~~~+ \bar{\chi}^I_\alpha \bar{\partial} \chi^I_\alpha +
i A^I_\mu \left( 
\bar{\chi}^I_1 \bar{\partial} X^\mu \chi^I_1 + \bar{\chi}^I_2 \bar{\partial} X^\mu \chi^I_2\right)\Bigr)
\end{eqnarray*}
where we have defined the coordinates on the torus~:
$z=\frac{1}{2}(\sigma^1+\tau \sigma^2), \ (\sigma^1, \sigma^2)\in [0, 1]^2$ and
the derivative $\partial = i(\bar{\tau} \partial_1 -
\partial_2)/\tau_2$. 

In this background, the $[1,1], [1,g_1], [1,g_2]$ and $[1,g_1g_2]$
contributions to the one-loop partition function read~:
\begin{eqnarray}
&&Z_{(1,1)}(v) = \frac{Q(\bar{\tau})}{8 \tau^4_2}
\frac{\sum_{a,b=0,1} 
\prod_{I=1}^{8}{\vartheta}^{2}\left[{}^{a}_{b}\right](v^I|\tau)}{\eta^{24}(\tau)}
\sum_{m^i, n^i \in {\mathbb Z}} \Gamma_{2,2}\ar{2 n^1 \;\; 2 n^2}{2 m^1 \; 2
m^2}(\tau, \bar{\tau}) \nonumber\\
&&Z_{(1,g_1)}(v) = \frac{Q(\bar{\tau})}{4 \tau^4_2}
\frac{
\prod_{I=1}^{8}{\vartheta}^{8}_3(v^I|\tau){\vartheta}^{8}_4(v^I|\tau)}{\eta^{24}(\tau)}
\sum_{m^i, n^i \in {\mathbb Z}} \Gamma_{2,2}\ar{2 n^1 \;\;\;\;\;\;\; 2 n^2}{2 m^1+1 \; 2
m^2}(\tau, \bar{\tau}) \nonumber\\
&&Z_{(1,g_2)}(v) = \frac{Q(\bar{\tau})}{4 \tau^4_2}
\frac{
\prod_{I=1}^{8}{\vartheta}_3(v^I|\tau){\vartheta}_4(v^I|\tau)}{\eta^{24}(\tau)}
\sum_{m^i, n^i \in {\mathbb Z}} \Gamma_{2,2}\ar{2 n^1 \;\; 2 n^2}{2 m^1 \; 2
m^2+1}(\tau, \bar{\tau}) \nonumber\\
&&Z_{(1,g_1g_2)}(v) = \frac{Q(\bar{\tau})}{4 \tau^4_2}
\frac{
\prod_{I=1}^{8}{\vartheta}_2(2v^I|2\tau)+\prod_{I=1}^{8}{\vartheta}_3(2v^I|2\tau)}{\eta^{8}(\tau)\eta^{8}(2\tau)}
\sum_{m^i, n^i \in {\mathbb Z}} \Gamma_{2,2}\ar{2 n^1 \;\;\;\;\;\;\; 2 n^2}{2 m^1+1 \; 2
m^2+1}(\tau, \bar{\tau}) \nonumber\\
\label{partfunct}
\end{eqnarray}
The contributions $[g_i, 1]$ and $[g_i, g_i]$  to the partition
function are 
obtained by doing the modular transformations $\tau
\rightarrow -1/\tau$ and $\tau \rightarrow (\tau-1)/\tau$ 
on $Z_{(1, g_i)}$. 

Taking the appropriate derivatives of this partition function, we can
extract the one-loop gauge thresholds~: 
\begin{eqnarray*}
{\cal I}^{\rm CHL}_{{\rm Tr}(F^4)} &=& \Delta^{\rm CHL}(T,U)\; t_8 {{\rm Tr}(F^4)},
\end{eqnarray*}
with
\begin{eqnarray}
\Delta^{\rm CHL}(T,U) &=& - \frac{{\cal N}}{3 \;2^5} \int_{\cal F} \frac{d^2
\tau}{\tau^2_2} \left(\Xi_{(1,1)} + (\Xi_{(1,g_1)} + \Xi_{(1,g_2)} +
\Xi_{(1,g_1 g_2)} 
+ {\rm orb.})\right)(\tau). \label{chltr1}
\end{eqnarray}
and
\begin{eqnarray*}
\Xi_{(1,1)}(\tau) &=&  
2\sum_{\alpha=2,3,4} \frac{\vartheta^{16}_\alpha}{\eta^{24}}
\left(\frac{\vartheta^{''''}_\alpha}{\vartheta_\alpha} - 3 \left(\frac{\vartheta^{''}_\alpha}{\vartheta_\alpha}\right)^2
\right) (\tau)
\sum_{m^i, n^i \in {\mathbb Z}} \Gamma_{2,2}\ar{2 n^1 \;\; 2 n^2}{2 m^1 \; 2
m^2}(\tau, \bar{\tau}), \nonumber\\
\Xi_{(1,g_1)}(\tau) &=&
2\sum_{\alpha=3,4} \frac{\vartheta^8_3\vartheta^8_4}{\eta^{24}}\left(
\frac{\vartheta^{''''}_\alpha}{\vartheta_\alpha} - 3 \left(\frac{\vartheta^{''}_\alpha}{\vartheta_\alpha}\right)^2
\right)(\tau)
\sum_{m^i, n^i \in {\mathbb Z}} 
\Gamma_{2,2}\ar{2 n^1 \;\;\;\;\;\;\; 2 n^2}{2 m^1+1 \; 2 m^2}(\tau,
\bar{\tau}), \nonumber\\ 
\Xi_{(1,g_2)}(\tau) &=&
2\sum_{\alpha=3,4} \frac{\vartheta^8_3\vartheta^8_4}{\eta^{24}}\left(
\frac{\vartheta^{''''}_\alpha}{\vartheta_\alpha} - 3 \left(\frac{\vartheta^{''}_\alpha}{\vartheta_\alpha}\right)^2
\right)(\tau)
\sum_{m^i, n^i \in {\mathbb Z}} \Gamma_{2,2}\ar{2 n^1 \;\; 2 n^2}{2 m^1 \; 2
m^2+1}(\tau, \bar{\tau}), \nonumber\\
\Xi_{(1,g_1g_2)}(\tau) &=&
\frac{1}{\eta^8(\tau)}\Bigl[\sum_{\alpha=2,3}\frac{\vartheta^8_\alpha}{\eta^8}
\left(
\frac{\vartheta^{''''}_\alpha}{\vartheta_\alpha} - 3 \left(\frac{\vartheta^{''}_\alpha}{\vartheta_\alpha}\right)^2
\right)\Bigr](2\tau)
\sum_{m^i, n^i \in {\mathbb Z}} \Gamma_{2,2}\ar{2 n^1 \;\;\;\;\;\;\; 2
n^2}{2 m^1+1 \; 2 
m^2+1}(\tau, \bar{\tau}) 
\end{eqnarray*}
Then, one can use the formula (\ref{derivatives}) and (\ref{doubling}) to
simplify these expressions. The elliptic functions cancel out and the
result is simply~:
\begin{eqnarray*}
\Xi_{(1,1)}(\tau) &=& {3\;2^6} \sum_{m^i, n^i \in {\mathbb Z}} \Gamma_{2,2}\ar{2 n^1 \;\; 2 n^2}{2 m^1 \; 2
m^2}(\tau, \bar{\tau}),  \nonumber\\
\Xi_{(1,g_1)}(\tau) &=&- 2^6 \sum_{m^i, n^i \in {\mathbb Z}} \Gamma_{2,2}\ar{2 n^1 \;\;\;\;\;\;\; 2 n^2}{2 m^1+1 \; 2
m^2}(\tau, \bar{\tau}), \nonumber\\
\Xi_{(1,g_2)}(\tau) &=& - 2^6
\sum_{m^i, n^i \in {\mathbb Z}} \Gamma_{2,2}\ar{2 n^1 \;\; 2 n^2}{2 m^1 \; 2
m^2+1}(\tau, \bar{\tau}) ,  \nonumber\\
\Xi_{(1,g_1g_2)}(\tau) &=& 2^6 
\sum_{m^i, n^i \in {\mathbb Z}} \Gamma_{2,2}\ar{2 n^1 \;\;\;\;\;\;\; 2
n^2}{2 m^1+1 \; 2 
m^2+1}(\tau, \bar{\tau})  
\end{eqnarray*}
Therefore, we can  group these contributions and rewrite the one-loop 
thresholds (\ref{chltr1}) as~: 
\begin{eqnarray*}
\Delta^{\rm CHL}(T,U) &=& -\frac{2{\cal N}}{3} \int_{\cal F} 
\frac{d^2\tau}{\tau^2_2} 
\sum_{m^i, n^i \in {\mathbb Z}} 
\Bigl( 
4 \Gamma_{2,2}\ar{2 n^1 \;\; 2 n^2}{2 m^1 \; 2 m^2}(\tau, \bar{\tau}) 
- 2 \Gamma_{2,2}\ar{n^1 \;\; 2 n^2}{m^1 \; 2 m^2}(\tau, \bar{\tau}) 
\nonumber\\
&&~~~~~~~~~~~~~~~~~~~~~~~
- 2 \Gamma_{2,2}\ar{2 n^1 \;\; n^2}{2 m^1 \; m^2}(\tau, \bar{\tau}) 
+ \Bigl(\Gamma_{2,2}\ar{2 n^1 \; 2 n^2}{m^1 \;\; m^2}(\tau, \bar{\tau})  
+{\rm orb.} \Bigr)\Bigr)
\end{eqnarray*}
which can be evaluated using the method of orbits. To calculate the
contribution of the last term, we use the technics described in
details and a more general context in the appendix of \cite{kop};
first, the lattice sum
$$\sum_{m^i, n^i 
\in 
{\mathbb Z}}\Gamma^{T, U}_{2,2}\ar{2 n^1 \; 2 n^2}{m^1 \;\; m^2}(\tau,
\bar{\tau}),$$ can be written as $$\frac{1}{2}\sum_{m^i, n^i \in
{\mathbb Z}}\Gamma^{2T, U}_{2,2}\ar{n^1 \;\; n^2}{m^1 \; m^2}(2\tau,
2\bar{\tau})$$ where we have reinstated explicitly the dependence on the
moduli of the torus. This term is integrated over the extended
fundamental domain ${\cal F}_2^-$ of the index 2 subgroup $\Gamma^-_2$
of $SL(2, {\mathbb Z})$, generated by $T$ and $ST^2S$. We can change
the integration variable to $\rho = 2\tau$~:
\begin{equation}
\int_{{\cal F}_2^-} \frac{d^2\tau}{\tau_2^2} \sum_{m^i, n^i \in
{\mathbb Z}}\Gamma^{T, U}_{2,2}\ar{2 n^1 \; 2 n^2}{m^1 \;\; m^2}(\tau,
\bar{\tau}) = \frac{1}{2} \int_{{\cal F}_2^+} \frac{d^2\rho}{\rho_2^2}\sum_{m^i, n^i \in
{\mathbb Z}}\Gamma^{2T, U}_{2,2}\ar{n^1 \;\; n^2}{m^1 \; m^2}(\rho,
\bar{\rho})
\end{equation}
and unfold the domain of integration into the fundamental domain of
$SL(2, {\mathbb Z})$
\begin{equation}
\int_{{\cal F}_2^-} \frac{d^2\tau}{\tau_2^2} \sum_{m^i, n^i \in
{\mathbb Z}}\Gamma^{T, U}_{2,2}\ar{2 n^1 \; 2 n^2}{m^1 \;\; m^2}(\tau,
\bar{\tau}) = \frac{3}{2} \int_{{\cal F}_2}
\frac{d^2\tau}{\tau_2^2}\sum_{m^i, n^i \in 
{\mathbb Z}}\Gamma^{2T, U}_{2,2}\ar{n^1 \;\; n^2}{m^1 \; m^2}(\tau,
\bar{\tau})
\end{equation}
Finally, reinstating the moduli dependence of the three first terms,
the threshold corrections read~:
\begin{eqnarray}
\Delta^{\rm CHL}(T,U) &=& -\frac{2{\cal N}}{3} \int_{\cal F} \frac{d^2
\tau}{\tau^2_2} \sum_{m^i, n^i \in {\mathbb Z}} 
\Bigl( 
 \Gamma^{4T,U}_{2,2}  - \Gamma^{2T,2U}_{2,2} - \Gamma^{2T,U/2}_{2,2} +
\frac{3}{2}\Gamma^{2T,U}_{2,2}  \Bigr)\ar{n^1 \; n^2}{m^1 m^2}(\tau,
\bar{\tau}) \nonumber \\
&=& \frac{{\cal N}}{3} \Bigl( 
2 {\rm log}\vert\eta(4T)\vert^4 
- {\rm log}\vert\eta(2T)\vert^4 
\nonumber \\
&&~~
+ 5 {\rm log}\vert\eta(U)\vert^4 
- 2 {\rm log}\vert\eta(2U)\vert^4 
- 2 {\rm log}\vert\eta(U/2)\vert^4 
\Bigr)  
\label{chlthres2} 
\end{eqnarray}

\section{Interpretation in type I
compactifications without vector structure}

\subsection{Type I compactification without vector structure, its
T-dual and D-instanton corrections}

The CHL string model is supposed to be S-dual to type I
string theory compactified on a torus without vector structure
\cite{bps,bianchi,witten}. This 
compactification has a discrete half integer NS-NS $B$ field flux on the
torus. Thanks to periodicity properties, we can always choose $B_{\rm 
NSNS}=\frac{1}{2}$ so the K\"ahler moduli of $T^2$ is 
\begin{eqnarray*}
T^{\rm
I}=\frac{1}{2}+iR^{{\rm I}^\prime}_1 
R^{{\rm I}^\prime}_2.
\end{eqnarray*}  
The mapping of the CHL moduli $T$ and $U$ to
type I moduli is  
identical to their standard heterotic and type I counterparts, namely
\begin{eqnarray*}
\tau^{{\rm I}} = B_{RR} + i {R^{\rm I}_1 R^{\rm I}_2}/{\lambda^{{\rm I}}_s} \qquad {\rm and} \qquad 
U^{\rm I}=i {R^{\rm I}_2}/{R^{\rm I}_1}.
\end{eqnarray*}  
The rank reduction of the group has a clear geometrical
interpretation if we perform a T-duality along one of the directions
of the torus, say $x^1$. This transformation exchanges the K\"ahler
and the complex structure of $T^2$. Therefore, we obtain a tilted
torus (fig. \ref{tilted}),
whose complex structure is
\begin{eqnarray*}
U^{{\rm I}^\prime}= \frac{1}{2} + i {R^{{\rm
I}}_1}{R^{{\rm I}}_2} = \frac{1}{2} + i \frac{R^{{\rm
I}^\prime}_2}{R^{{\rm I}^\prime}_1} . 
\end{eqnarray*}
\FIGURE{\begin{picture}(380,190)(0,0)
\put(50,0){\epsfig{file=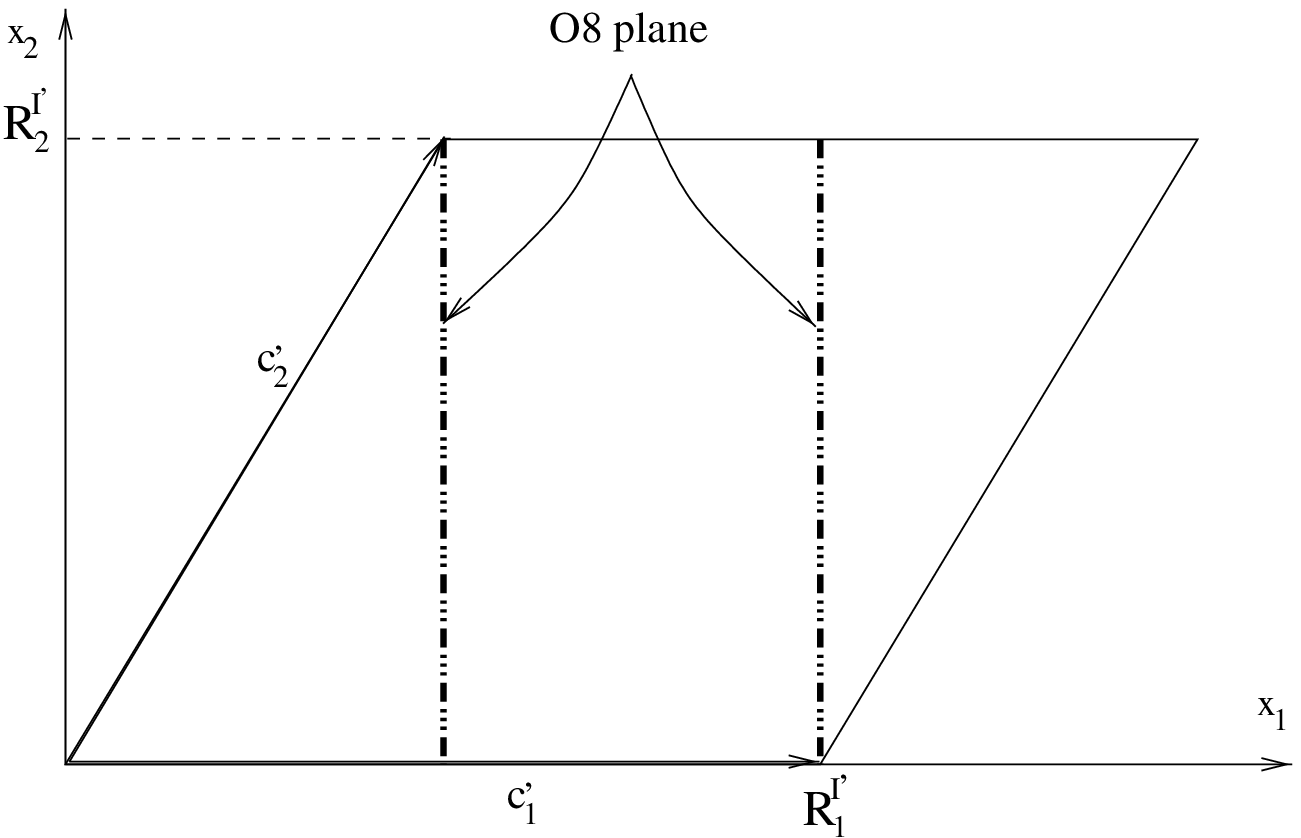,height=6.5cm}}
\end{picture}
\caption{Type I' string on a tilted torus.}
\label{tilted}}

After T-duality, the orientifold projection is mapped to $\Omega {\cal
R}_1$ and the O9-plane is transformed into O8-planes. However, contrary to
the compactification on a torus with vector structure, there is only one
O8-plane, wrapped on the cycle $2c^\prime_2-c^\prime_1$ of the torus as one
can see on the figure \ref{tilted}. To cancel the unphysical R-R flux
induced by the presence of this plane, one introduces D8-branes which,
as it is easy to verify using the standard T-duality transformations
on the boundary conditions for open strings ending on the D-branes,
are oriented along the same cycle and, hence, wrapped twice the
torus. Therefore, compared to the standard case, only half of these
D8-branes are needed and the rank of the gauge group is divided by
two. Rather than to have two systems of [1 O8-plane+8 D8-branes]
wrapped once on a square torus, we have only one system wrapped twice
on the skew torus with angle $\pi/3$. 

This interpretation is consistent with the spacetime anomaly
cancellation. Indeed, we know that gravitational anomalies are
canceled due to the presence of Wess-Zumino gravitational couplings
in the actions of the D-branes and of the orientifold planes
\cite{ghm}. Since 
the D8-branes and the O8-planes are wrapped twice on the torus, they
contribute exactly twice and cancel the same amount of anomaly as
two O8-planes and sixteen D8-branes of the standard type I'
compactification.  

We are now in position to interpret the CHL thresholds
(\ref{chlthres2}) in this type I picture. Similarly to the first
section, the 
$T$-dependent part corresponds to a tree-level contribution coming from
the expansion of the Born-Infeld action and of D-instanton
contributions due to euclidian D0-branes~:
\begin{eqnarray*}
\Delta^{\rm CHL}(T) &=& \frac{{\cal N}}{3} 
\left(
- 2\pi T_2 + 4 {\rm Re} \sum_N\sum_{m|N}
\frac{1}{m}{\rm e}^{8i \pi N T} - 2{\rm Re}
\sum_N \sum_{m|N} \frac{1}{m}{\rm e}^{4i \pi N T}  
\right)
\label{chlnp}
\end{eqnarray*}
One the type I' side, the parameter $T$ becomes $A_R + i R^{\rm
I'}_2/g^{\rm I'}_2 $. It
corresponds to the effective action of an euclidian D-particle whose
world-line describes the cycle $2c^\prime_2 - c^\prime_1$. We observe that these
non-perturbative corrections are exactly the same as (\ref{hetnp}), except
that $T$ has been replaced by $2T$ as one expects from the double
wrapping required for closure of the D-particle world-line on the
tilted torus.  
It would be very interesting to understand this result from a
non-perturbative calculation, such as in the matrix model perspective
suggested by  
\cite{gut1}. Indeed, the quantum mechanics which describes the
D0-branes on top of the [O8-plane + 8 D8-branes] system is the same,
only the boundary conditions on the fields are modified.

\subsection{One-loop ${\rm Tr}(F^4)$ couplings}

To test the S-duality relation between CHL string and type I
compactification on a torus without vector structure, we can repeat
the one-loop calculation made in section two in the 
context of type I 
string with gauge group $SO(16)\times SO(16)$. In type I
compactifications without vector structure, the presence of a
background  field leads to the following modified open
string partition 
functions on the annulus and on the M{\oe}bius strip~:
\begin{eqnarray*}
{\cal A}\left(B\right) &=& \frac{i V^{(8)}}{3\;2^{13}\pi^4} 
\int^\infty_0 \frac{dt}{t}
\frac{1}{\left(2\pi^2 t\right)^4}
\frac{\vartheta_1^\prime\left(0|\frac{it}{2}\right)}
{\vartheta_1\left(\frac{i\epsilon t}{2}|\frac{it}{2}\right)}
\frac{i}{2}{\left(q_i+q_j\right)Bt} 
\sum_{m_i \in {\mathbb Z}} 
\Gamma_2[m_1 \; m_2]\left(\frac{it}{2}\right)
 \nonumber\\
&&~~~~~~~\times
\sum_{a,b=0,1}
\frac{1}{2}\left(-\right)^{a+b+ab} 
\frac{\vartheta^3\left[{}^{a}_{b}\right]\left(0|\frac{it}{2}\right)
\vartheta\left[{}^{a}_{b}\right]\left(\frac{i\epsilon
t}{2}|\frac{it}{2}\right)}{\eta^{12}\left(\frac{it}{2}\right)}, \nonumber\\
{\cal M}\left(B\right) &=& 
-\frac{i V^{(8)}}{3\;2^{13}\pi^4} 
\int^\infty_0 \frac{dt}{t}
\frac{1}{\left(2\pi^2 t\right)^4}
\frac{\vartheta_1^\prime\left(0|\frac{it+1}{2}\right)}
{\vartheta_1\left(\frac{i\epsilon t}{2}|\frac{it+1}{2}\right)}
\left(i q_iBt\right)
\sum_{{}^{\epsilon_i =0,1}_{m_i \in 2{\mathbb
Z}+\epsilon_i}} \left(-\right)^{\epsilon^1 \epsilon^2} \Gamma_2[m_1 \;
m_2]\left(\frac{it}{2}\right)  
\nonumber\\
&&~~~~~~~\times 
\sum_{a,b=0,1}
\frac{1}{2}
\left(-\right)^{a+b+ab} 
\frac{\vartheta^3\left[{}^{a}_{b}\right]\left(0|\frac{it+1}{2}\right)
\vartheta\left[{}^{a}_{b}\right]\left(\frac{i\epsilon
t}{2}|\frac{it+1}{2}\right)}{\eta^{12}\left(\frac{it+1}{2}\right)} 
\end{eqnarray*}
whose quartic expansions in $B$ give
\begin{eqnarray*}
{\cal A}|_{B^4} &=& -\frac{V^{(8)}}{3\;2^{12}\pi^4} 
(q_i + q_j)^4 B^4
\int_0^\infty \frac{dt}{t} 
\sum_{m_i \in {\mathbb Z}} 
\Gamma_2[m_1 \; m_2]\left(\frac{it}{2}\right), \nonumber\\
{\cal M}|_{B^4} &=&
\frac{V^{(8)}}{3 \;2^{12}\pi^4} 
(2q_i)^4 B^4
\int_0^\infty \frac{dt}{t} 
\sum_{{}^{\epsilon_i =0,1}_{m_i \in {\mathbb
Z}}} \left(-\right)^{\epsilon_1 \epsilon_2} \Gamma_2[2m_1+\epsilon_1 \;\;
2m_2+\epsilon_2]\left(\frac{it}{2}\right)  
\end{eqnarray*}
From these amplitudes, we can extract the threshold corrections to the
${\rm Tr}(F^4)$ terms~:
\begin{eqnarray*}
{\cal I}^{\rm 1-loop}_{{\rm Tr}(F^4)} = \Delta^{\rm I}(U)\; t_8 {{\rm
Tr}(F^4)} 
\end{eqnarray*}
with
\begin{eqnarray*}
\Delta^{\rm I}(U) &=& -\frac{V^{(8)}}{3\;2^{12}\pi^4} 
\int_0^\infty \frac{dt}{t} 
\Bigl(
2 \times 16 \sum_{m_i \in {\mathbb Z}} \Gamma_2[m_1 \;
m_2] \nonumber \\
&&~~~~~~~~~~~~~~~~~~~~~~~~~- 
2^4 \sum_{{}^{\epsilon_i =0,1}_{m_i \in {\mathbb Z}}} 
\left(-\right)^{\epsilon_1 \epsilon_2} 
\Gamma_2[2m_1+\epsilon_1 \;\;
2m_2+\epsilon_2] 
\Bigr)
\left(\frac{it}{2}\right)  \nonumber \\
&=& \frac{V^{(8)}}{3\; 2^{8}\pi^4}  
\Bigl( 
5{\rm log}{|\eta(U)|}^4 - 2 {\rm log}{|\eta(2U)|}^4 - 2 {\rm
log}{|\eta(U/2)|}^4  
\Bigr)
\end{eqnarray*}
which is the complex structure dependent part of the CHL correction
$\Delta^{\rm CHL}(U)$. Therefore, this calculation provides a
quantitative test 
of the S-duality which relates the CHL string to type I
compactification without vector structure.

\section{Conclusion}

In this note, we have investigated half-BPS saturated couplings in the
effective action of the CHL string and of its type I dual. Comparing part of
the threshold corrections to the heterotic ${\rm Tr}(F^4)$ terms to one-loop
corrections on the type I side has provided a quantitative test of the
duality between the CHL string and the type I string compactified on a
torus without vector structure. 

This calculation also predicts the existence of D-instanton
corrections due to D-strings wrapping the torus of the type I string
theory. After one T-duality, these corrections can be attributed to
D-particles whose euclidean world-lines wrap twice the dual skew torus. A
striking feature of these instantonic corrections is that there are
identical (up to an obvious factor related to the length of the
world-line of the D-particle) to the D-instanton corrections
which appear in another 
compactification, namely type I' string with two orientifold
8-planes and sixteen D8-branes, with gauge group is
$SO(16)\times SO(16)$.  From the 11-dimensional point-of-view,
these two type I compactifications can be seen respectively as ${\cal
M}$-theory 
compactified on a M{\oe}bius strip and on an annulus, the gauge groups
$SO(16)$ living on the boundaries of these open Riemann
surfaces.  The co\"{\i}ncidence that we have observed in this article
seems to indicate that 
instanton corrections to the half-BPS saturated gauge couplings
are only a ``boundary effect''. It would be 
very interesting to recover these corrections from a 
non-perturbative description such as the matrix model describing
D0-branes on top of one O8-plane and 8 D8-branes. 
We leave this question for future work.

%%%%%%%%%%%%%%%%%%%%%%%%%%%%%%%%%%%%%%%%%%%%%%%%%%%%%%%%%%%%%%%%%%%%%%%%%%%%%%%%%%%%%%%%%%%%%%%%%

\ 

\

\noindent{\bf Acknowledgements} 

We have benefited from discussions with C. Bachas, M. Bianchi,
M. Green, E. Kiritsis and
A. Sagnotti.  This work is supported by PPARC.

\

\appendix{\noindent\Large {\bf{Appendix}}} 
\label{appendix} 

\section{Theta function identities}

\centerline{\bf  Definition}
\begin{equation}
\vartheta\left[{}^{a}_{b}\right](v|\tau) = \sum_{n\in \mathbb{Z}}
q^{(n-\frac{a}{2})^2} e^{2i\pi\left(v-\frac{b}{2}\right)\left(n-\frac{a}{2}\right)}
\end{equation}

\centerline{\bf  Jacobi identity}
\begin{eqnarray}
&&\sum_{a,b=0,1} (-)^{a+b+ab}
\vartheta\left[{}^{a}_{b}\right](v_1|\tau)\vartheta\left[{}^{a}_{b}\right](v_2|\tau)\vartheta\left[{}^{a}_{b}\right](v_3|\tau)\vartheta\left[{}^{a}_{b}\right](v_4|\tau)
\nonumber \\
&&~~~~~~~~~~~~~~~~~~~~~~~~~~~~~~~= -2
\vartheta_1(v^\prime_1|\tau)\vartheta_1(v^\prime_2|\tau)\vartheta_1(v^\prime_3|\tau)\vartheta_1(v^\prime_4|\tau)
\label{jacobi}
\end{eqnarray}
with
\begin{eqnarray*}
&&v^\prime_1 = \frac{1}{2} \left(-v_1 + v_2 + v_3 + v_4  \right), ~~~~ 
v^\prime_2 = \frac{1}{2} \left(v_1 - v_2 + v_3 + v_4  \right), \\
&&v^\prime_3 = \frac{1}{2} \left(v_1 + v_2 - v_3 + v_4  \right), ~~~~ 
v^\prime_4 = \frac{1}{2} \left(v_1 + v_2 + v_3 - v_4  \right),
\end{eqnarray*}
\centerline{\bf  Derivatives of theta functions}
We use a prime to denote the derivative of a theta function with
respect to $v$. The Jacobi theta functions verify the relation 
\begin{equation}
\left(i\pi \partial^{(2)}_v - \partial_\tau
\right)\vartheta\left[{}^{a}_{b}\right](v|\tau) = 0.
\end{equation}
The relevant formula for the calculations of this article are given by
\begin{eqnarray}
&&\frac{\vartheta_2^{''}}{\vartheta_2}=\frac{1}{12}\left(\hat{E}_2+\vartheta_3^4+\vartheta_4^4\right),\qquad
\frac{\vartheta_2^{''''}}{\vartheta_2}-3\left(\frac{\vartheta_2^{''}}{\vartheta_2}\right)^2=-\frac{1}{8}\vartheta_3^4\vartheta_4^4,\nonumber
\\
&&\frac{\vartheta_3^{''}}{\vartheta_3}=\frac{1}{12}\left(\hat{E}_2+\vartheta_2^4-\vartheta_4^4\right),\qquad
\frac{\vartheta_3^{''''}}{\vartheta_3}-3\left(\frac{\vartheta_3^{''}}{\vartheta_3}\right)^2=\frac{1}{8}\vartheta_2^4\vartheta_4^4,\nonumber
\\
&&\frac{\vartheta_4^{''}}{\vartheta_4}=\frac{1}{12}\left(\hat{E}_2-\vartheta_2^4-\vartheta_3^4\right),\qquad
\frac{\vartheta_4^{''''}}{\vartheta_4}-3\left(\frac{\vartheta_4^{''}}{\vartheta_4}\right)^2=-\frac{1}{8}\vartheta_2^4\vartheta_3^4.
\label{derivatives}
\end{eqnarray}
\centerline{\bf  Doubling formula}
\begin{eqnarray}
\eta(2\tau)\vartheta_4(2\tau) = \eta^2(\tau).
\label{doubling}
\end{eqnarray}

\ 

\section{World-sheets integrations}

The purpose of this appendix is to perform the relevant integrations
over the moduli which parameterize the one-loop world-sheets of the
closed and open strings. 

\ 

\centerline{\bf  Torus}

For the toroidal partition function, we have to calculate a modular
integral over the fundamental domain of $SL(2, {\mathbb Z})$. To do
this, one unfolds this domain using the method of orbits
\cite{dkl}. This technics leads to the result~:
\begin{eqnarray}
\int_{\cal F}\frac{d^2\tau}{\tau_2^2} \sum_{m^i, n^i \in {\mathbb Z}} 
\Gamma^{T,U}_{2,2}\ar{n^1 \; n^2}{m^1 m^2}(\tau,\bar{\tau}) = 
- {\rm log}\left(|\eta(T)|^4|\eta(U)|^4\right). 
\label{clint}
\end{eqnarray}
Actually, the integral has a logarithmic infrared divergence due to
massless 
string modes circulating into the loop that has not been written in
the above formula. 

\ 

\centerline{\bf  Annulus and M{\oe}bius strip}

The calculation of open string amplitudes is more subtle since there
is no modular invariance to cut off the quadratic ultraviolet
divergence of the 
integrals over the open string world-sheets. However, when tadpoles are
canceled, the ultraviolet divergences of the annulus and of the M{\oe}bius
strip cancel each other. This is always the case for the models
considered in this paper. For each diagram, the divergence is given by
the zero windings sector after a double Poisson resummation of the
open-strings Kaluza-Klein momenta to the closed channel windings.  
 
Therefore, the formula needed in this article is~:
\begin{eqnarray}
\int_0^\infty \frac{dt}{t} 
&&\left(
\sum_{{m_i \in {\mathbb Z}}}
\Gamma_2[m_1 \;m_2]\left(\frac{it}{2}\right) - \frac{\pi T_2}{t}
\right)
= - 
{\rm log}\vert\eta(U)\vert^4 
\label{opint}
\end{eqnarray}
where we have explicitly subtracted the quadratic ultraviolet
divergence part. These terms cancel out each other after adding the
annulus and M{\oe}bius strip amplitudes. Moreover, we have also
overlooked the logarithmic infrared divergence due to massless
open-string states running into the loop. Finally, integrals of
lattice sums with shifts of $1/2$ on the momenta can be obtained from
this formula as sum of terms with different complex structure
dependences~:
\begin{eqnarray}
&&\int_0^\infty \frac{dt}{t} 
\sum_{{}^{m_1 \in {\mathbb Z}+\alpha_1}_{m_2 \in {\mathbb Z}+\alpha_2}}
\Gamma_2[m_1 \;m_2]\left(\frac{it}{2}\right) 
\nonumber \\
&&~~~~~= 
{\rm log}\vert\eta(U)\vert^4 -
{\rm log}\vert\eta(U/2)\vert^4
\mbox{~~~for~~~} 
(\alpha_1, \alpha_2)=(\frac{1}{2},0),
\nonumber \\
&&~~~~~=  
{\rm log}\vert\eta(U)\vert^4 - 
{\rm log}\vert\eta(2U)\vert^4
\mbox{~~~for~~~} 
(\alpha_1, \alpha_2)=(0,\frac{1}{2}),
\nonumber \\
&&~~~~~= 
{\rm log}\vert\eta(U/2)\vert^4 
+ {\rm log}\vert\eta(2U)\vert^4 
- 
2{\rm log}\vert\eta(U)\vert^4
\mbox{~~~for~~~} 
(\alpha_1, \alpha_2)=(\frac{1}{2},\frac{1}{2}).
\label{opint2}
\end{eqnarray}

%%%%%%%%%%%%%%%%%%%%%%%%%%%%%%%%%%%%%%%%%%%%%%%%%%%%%%%%%%%%%%%%%%%%%%%%%

\begin{thebibliography}{99}


\bibitem{w1} 
E. Witten, {\em String Theory Dynamics In Various Dimensions}, 
Nucl. Phys.   {\bf B443} (1995) 85, \hepth{9503124}.

\bibitem{pw} 
J. Polchinski and E. Witten, {\em Evidence for Heterotic - Type I
String Duality},  
Nucl. Phys. {\bf B460} (1996) 525, \hepth{9510169}.

\bibitem{bk} 
C. Bachas and E. Kiritsis, {\em $F^4$ Terms in N=4 String Vacua}, 
Nucl. Phys. Proc. Suppl.  {\bf 55B} (1997) 194, \hepth{9611205}.

\bibitem{bfkov} 
C. Bachas, C. Fabre, E. Kiritsis, N.A. Obers and P. Vanhove, {\em
Heterotic / type I duality and D-brane instantons},  
Nucl. Phys.  {\bf B509} (1998) 33, \hepth{9707126}.

\bibitem{ko} 
E. Kiritsis and N.A. Obers, {\em Heterotic/Type-I Duality in $D<10$
Dimensions, Threshold Corrections and D-Instantons},  
JHEP  {\bf  9710} (1997) 004, \hepth{9709058}.

\bibitem{b1} 
C. Bachas, {\em Heterotic versus Type I}, 
Nucl. Phys. Proc. Suppl.  {\bf 68} (1998) 348, \hepth{9710102}.

\bibitem{fs} 
K. Foerger and S. Stieberger, {\em Higher Derivative Couplings and
Heterotic-Type I Duality in Eight Dimensions},  
 Nucl. Phys. {\bf B559} (1999) 277, \hepth{9901020}.

\bibitem{gut1} 
M. Gutperle, {\em A note on heterotic/type I' duality and D0 brane
quantum mechanics},  
 JHEP  {\bf 9905} (1999) 007, \hepth{9903010}.

\bibitem{ls} 
W. Lerche and S. Stieberger, {\em Prepotential, Mirror Map and
F-Theory on K3},  
Adv. Theor. Math. Phys.   {\bf 2} (1998) 1105, \hepth{9804176}; Erratum-ibid. {\bf 3} (1999) 1199.

\bibitem{lsw} 
W. Lerche, S. Stieberger and N.P. Warner, {\em Quartic Gauge Couplings from K3 Geometry}, 
 Adv. Theor. Math. Phys. {\bf  3} (1999) 1575, \hepth{9811228}.

\bibitem{kop} 
E. Kiritsis, N.A. Obers and B. Pioline, {\em Heterotic/Type II Triality and Instantons on $K_3$}, 
JHEP  {\bf 0001} (2000) 029, \hepth{0001083}.

\bibitem{df} 
U. Danielsson and G. Ferretti, {\em The Heterotic Life of the D-particle}, 
Int. J. Mod. Phys. {\bf A12} (1997) 4581, \hepth{9610082}.

\bibitem{bss} 
T. Banks, N. Seiberg and E. Silverstein, {\em Zero and One-dimensional
Probes with N=8 Supersymmetry},  
Phys. Lett. {\bf B401} (1997) 30, \hepth{9703052}.

\bibitem{bgl} 
O. Bergman, M.R. Gaberdiel and G. Lifschytz, {\em String Creation and
Heterotic-Type I' Duality},  
Nucl. Phys.  {\bf B524} (1998) 524, \hepth{9711098}.

\bibitem{bgs} 
C. Bachas, M. B. Green and  A. Schwimmer, {\em (8,0) Quantum mechanics
and symmetry enhancement in type I' superstrings},  
 JHEP  {\bf 9801} (1998) 006, \hepth{9712086}.

\bibitem{gg1} 
M.B. Green and M. Gutperle, {\em Effects of D-instantons}, 
Nucl. Phys.  {\bf B498} (1997) 195, \hepth{9701093}.

\bibitem{gg2} 
M.B. Green and M. Gutperle, {\em D-instanton partition functions}, 
Phys. Rev.  {\bf D58} (1998) 046007, \hepth{9804123}.

\bibitem{mns} 
G. Moore, N. Nekrasov and S. Shatashvili, {\em D-particle bound states
and generalized instantons},  
 Commun. Math. Phys. {\bf 209} (2000) 77, \hepth{9803265}.

\bibitem{kv} 
I. Kostov and P. Vanhove, {\em Matrix String Partition Functions}, 
 Phys. Lett. {\bf B444} (1998) 196, \hepth{9809130}.

\bibitem{bfss} 
T. Banks, W. Fischler, S.H. Shenker and L. Susskind, {\em M Theory As
A Matrix Model: A Conjecture},  
 Phys. Rev.  {\bf D55} (1997) 5112, \hepth{9610043}.

\bibitem{bbg} 
C.P. Bachas, P. Bain and M.B. Green, {\em Curvature terms in D-brane
actions and their M-theory origin},  
 JHEP  {\bf 9905} (1999) 011, \hepth{9903210}.

\bibitem{bps} 
M. Bianchi, G. Pradisi and A. Sagnotti, {\em Toroidal compactification
and symmetry breaking in open string theory}, 
Nucl. Phys.  {\bf B376} (1992) 365.

\bibitem{bianchi} 
M. Bianchi, {\em A Note on Toroidal Compactifications of the Type I Superstring and Other Superstring Vacuum Configurations with 16 Supercharges}, 
Nucl. Phys.  {\bf B528} (1998) 73, \hepth{9711201}.

\bibitem{witten} 
E. Witten, {\em Toroidal Compactification Without Vector Structure}, 
JHEP {\bf 9802} (1998) 006, \hepth{9712028}.

\bibitem{chl} 
S. Chaudhuri, G. Hockney and J. D. Lykken, {\em Maximally
Supersymmetric String Theories in $D<10$},  
 Phys. Rev. Lett.  {\bf 75} (1995) 2264, \hepth{9505054}.

\bibitem{dp} 
A. Dabholkar and J. Park, {\em Strings on Orientifolds}, 
 Nucl. Phys. {\bf B477} (1996) 701, \hepth{9604178}.

\bibitem{ejm} 
J. Ellis, P. Jetzer and  L. Mizrachi, {\em One-loop string corrections
to the effective field theory}, 
Nucl. Phys.  {\bf B303} (1988) 1.

\bibitem{dkl} 
L. Dixon, V. Kaplunovsky and J. Louis, {\em
Moduli dependence of string loop corrections to gauge coupling constants}, 
Nucl. Phys  {\bf B355} (1991) 64.

\bibitem{bp} 
C. Bachas and M. Porrati, {\em Pair Creation of Open Strings in an
Electric Field},  
Phys. Lett. {\bf B296} (1992) 77, \hepth{9209032}.

\bibitem{bf} 
C. Bachas and C. Fabre, {\em Threshold effects in open-string theory}, 
Nucl. Phys.  {\bf B476} (1996) 418, \hepth{9605028}.


\bibitem{bgmn} 
M. Bianchi, E. Gava, J. F. Morales and K. S. Narain, {\em D-strings in
unconventional type I vacuum configurations},  
Nucl. Phys.  {\bf B547} (1999) 96, \hepth{9811013}.

\bibitem{ghm} 
M. Green, J. Harvey and G. Moore, {\em I-Brane Inflow and Anomalous
Couplings on D-Branes},  
Class. Quant. Grav.  {\bf 14} (1997) , \hepth{9605033}.



\end{thebibliography}
\end{document}